# Bond Percolation on Multiplex Networks

A. Hackett,[1] D. Cellai,[1,2] S. Gómez,[3] A. Arenas,[3] and J. P. Gleeson[1]

[1]*MACSI, Department of Mathematics and Statistics, University of Limerick, Limerick, Ireland*
[2]*Idiro Analytics, Clarendon House, 39 Clarendon Street, Dublin 2, Ireland*
[3]*Departament d'Enginyeria Informàtica i Matemàtiques, Universitat Rovira i Virgili,
43007 Tarragona, Spain*

We present an analytical approach for bond percolation on multiplex networks and use it to determine the expected size of the giant connected component and the value of the critical bond occupation probability in these networks. We advocate the relevance of these tools to the modeling of multilayer robustness and contribute to the debate on whether any benefit is to be yielded from studying a full multiplex structure as opposed to its monoplex projection, especially in the seemingly irrelevant case of a bond occupation probability that does not depend on the layer. Although we find that in many cases the predictions of our theory for multiplex networks coincide with previously derived results for monoplex networks, we also uncover the remarkable result that for a certain class of multiplex networks, well described by our theory, new critical phenomena occur as multiple percolation phase transitions are present. We provide an instance of this phenomenon in a multiplex network constructed from London rail and European air transportation data sets.

## I. INTRODUCTION

In recent years, there has been a heightened interest within the network science community in the properties of multilayer networks [1,2]. In this regard, after the seminal work of Leicht and D'Souza [3], an extensive body of literature has emerged concerning the robustness of various subclasses of multilayer networks, most prominently interdependent networks [4–14], as well as interconnected networks [3,15–18], networks of networks [19,20], and multiplex networks [21–30].

Significant progress has been made in understanding the percolation properties of multilayer networks, and some surprising features have been uncovered [4,5,8,19]. For example, dependency links can have a dramatic impact on cascading failure events, and as a consequence, critical phenomena in a network formed of interdependent networks can be very different from those observed in monoplex networks. In some instances, this can mean that a network of networks as a whole will be more fragile than its constituent parts taken in isolation [8,10,31].

Ostensibly, the questions that are of most interest with regard to multilayer network robustness are well established. Except for Refs. [22,26], the aforementioned literature deals almost exclusively with site percolation and variants thereof. Moreover, by far the most common measure of robustness used is the expected size of the mutually connected giant component (MCGC). However, other existing ways of tackling the question of robustness should not be neglected. In particular, bond percolation and the expected size of the giant connected component (GCC) may be especially relevant to the study of the robustness of multiplex networks [32].

A multiplex network is a type of multilayer structure in which a set of $N$ nodes are interconnected by $M$ different sets of edges, where each set of edges exists in a unique layer and the same set of nodes is replicated across all layers. In the particular case in which there are no interlayer links (or when they can be discarded), a multiplex network reduces to an edge-colored graph in which the color of an edge corresponds to the set to which it belongs. Thus, layers (colors) can be used to represent distinct kinds of interactions. For example, the network of passenger airline routes within Europe can be represented as a multiplex network with nodes corresponding to airports, edges to routes, and each layer to the airline operating on a unique subset of these routes.

From a modeling perspective, site percolation on such a network, which begins with the removal of a fraction $1 - p$ of its nodes and all adjoining edges (where $p$ is termed the site occupation probability), corresponds to the removal of the same fraction of airports and all adjoining routes. This in turn can lead to the shutdown of airports that were once connected to those removed, and so on from airport to airport. When this propagation of failures ceases, the MCGC, if it exists, corresponds to the remaining component of the network that contains pairs of airports that are connected by all airlines, i.e., by routes of every color. By contrast, in bond percolation, a fraction $1 - p$ of edges are removed (where $p$ is now termed the bond occupation probability) corresponding to the removal by a set of airlines of their services to a set of airports. At the end of the propagation of failures instigated by this initial removal, the GCC, if it exists, corresponds to the remaining component of the network that contains pairs of airports that are connected by one or more airlines, i.e., by a route of any color. We contend that from the point of view of passengers navigating their way through the network, the GCC may be a more pertinent measure of robustness as their primary concern will generally be to get to their destination irrespective of the airline that gets them there. It also appears that a more likely cause of disruption might be the closure of routes not airports. Similar arguments can be given for studying bond percolation and the GCC on other varieties of multiplex transportation or communication networks.

However, a valid question may be raised as to whether it is necessary to consider the full multiplex structure in order to tackle such a problem [33]. After all, one could simply project a multiplex network to a monoplex network by ignoring the colors of edges and aggregating the layers. One would then calculate the expected GCC size in the type of network for which it was originally defined. In this paper, we address this question by revealing the structural constraints under which a multiplex network and its projection give the same results for the expected size of the GCC and the value of the critical bond occupation probability $p_c$. We find that a particular class of multiplex networks for which the results differ also manifests multiple percolation phase transitions reminiscent of those observed in Refs. [34] and [35] for clustered and modular networks, respectively.

The remainder of this paper is structured as follows. In Sec. II, we present our analytical approach for bond percolation on multiplex networks and show how to calculate the expected GCC size and $p_c$ in the edge-colored and projected versions of said networks, respectively. The results for each version together constitute two separate but overlapping theories. In Sec. III, we outline several cases in which these two theories coincide. Section IV describes the construction of a particular class of multiplex networks for which the theories differ dramatically in their predictions as outlined above. In Sec. V, we show that a multiplex network constructed from London rail [25] and European air [36] transportation data sets exhibits percolation behavior similar to that of the multiplex networks of Sec. IV. We conclude in Sec. VI.

## II. BOND PERCOLATION ANALYSIS

The fundamental property that we use to describe a multiplex network is its multidegree distribution $P_{\vec{k}}$, which gives the probability that a randomly chosen node in the network has multidegree vector $\vec{k} = (k_1, ..., k_\alpha, ..., k_M)$, where $k_\alpha$ is the degree of the node in layer $\alpha$, i.e., the number of edges of type $\alpha$ incident on the node. Let $K$ be the aggregated degree of the node over all layers. Then, the degree distribution of the projected network $\bar{P}_K$ is given by summing $P_{\vec{k}}$ over all multidegrees that sum to $K$:

$$\bar{P}_K = \sum_{\substack{\vec{k} \\ \sum_\alpha k_\alpha = K}} P_{\vec{k}}. \qquad (1)$$

We want to calculate the expected size of the GCC after a process in which each edge of the network is occupied with probability $p$ irrespective of its color. This is equivalent to calculating the probability $S$ that a randomly chosen node is in the GCC. As each layer of our multiplex is created by the configuration model and is of order $N \to \infty$, clustering is negligible [37], and we can approximate the network as a tree with a randomly chosen node $A$ as its root. Let the word *active* signify that a node is part of the GCC and *inactive* that it is not. Then, the value of $S$ is given by calculating the probability of activation of $A$ after a process of level-by-level activations from child nodes on one level to their parents on the next level closest to $A$.

Let $q_\alpha$ be the probability that a randomly selected node connected to its parent by an edge of type $\alpha$ is active given that its parent is not; then, for all $\alpha = 1, ..., M$, the following equation holds:

$$q_\alpha = 1 - \sum_{\vec{k}} \frac{k_\alpha}{\langle k_\alpha \rangle} P_{\vec{k}} (1 - pq_\alpha)^{k_\alpha - 1} \prod_{\beta \neq \alpha} (1 - pq_\beta)^{k_\beta}. \qquad (2)$$

The term $(k_\alpha/\langle k_\alpha \rangle) P_{\vec{k}}$ (where $\langle \cdot \rangle$ denotes the mean) is the probability that the node at the end of an edge of type $\alpha$ has multidegree $P_{\vec{k}}$ and the remaining terms in the summation express the probability that the children of this node are all inactive. Solving Eq. (2) for $q_\alpha$, we find $q_\alpha^*$, the probability that a child of $A$ connected to it by an edge of type $\alpha$ is active. We then have the following equation:

$$S = 1 - \sum_{\vec{k}} P_{\vec{k}} \prod_{\alpha=1}^{M} (1 - pq_\alpha^*)^{k_\alpha}. \qquad (3)$$

The differences between Eqs. (3) and (2) follow from the fact that the root has no parent.

Let us label the right-hand side of Eq. (2) as $Q_\alpha \equiv Q_\alpha(\vec{q})$. To find an expression for $p_c$, we must linearize $Q_\alpha$ and examine the eigenvalues of the matrix

$$\left|\left(\frac{\partial Q_\alpha}{\partial q_\beta}\right) - (1+\lambda)\mathbb{I}\right| = 0, \qquad (4)$$

where

$$\left.\frac{\partial Q_\alpha}{\partial q_\beta}\right|_{\bar{q}=0} = p\left(\frac{\langle k_\alpha k_\beta\rangle}{\langle k_\alpha\rangle} - \delta_{\alpha\beta}\right). \qquad (5)$$

The expected size of the GCC is nonzero when the largest eigenvalue of this matrix is positive, $\lambda_{\max} > 0$. One can define the matrix

$$B_{\alpha\beta} = \frac{\langle k_\alpha k_\beta\rangle}{\langle k_\alpha\rangle}, \qquad (6)$$

and it is easy to see that the largest eigenvalue $\mu_{\max}$ of **B** is related to $\lambda_{\max}$ as $\lambda_{\max} = p\mu_{\max} - p - 1$. Therefore, the critical bond occupation probability in the edge-colored network can be expressed as

$$p_c = \frac{1}{\mu_{\max} - 1}. \qquad (7)$$

The authors of Refs. [38,39] have applied a similar approach to find the critical bond occupation probability in a class of coupled networks, different from the multiplex networks of this paper, which they used to study overlaying social-physical networks.

Next, suppose that we ignore the colors of edges and aggregate the layers of the multiplex network, thereby obtaining its monoplex projection. By similar arguments as before, the probability that a randomly selected node in the projected network is active given that its parent is inactive is given by

$$\bar{q} = 1 - \sum_K \frac{K}{\langle K\rangle}\bar{P}_K(1 - p\bar{q})^{K-1}, \qquad (8)$$

and the expected size of the GCC is

$$\bar{S} = 1 - \sum_K \bar{P}_K(1 - p\bar{q}^*)^K, \qquad (9)$$

where $\bar{q}^*$ is found by solving Eq. (8) for $\bar{q}$. This result was previously derived in Ref. [40].

The critical bond occupation probability in the projected network may be obtained by labeling the right-hand side of Eq. (9) as $\bar{Q} \equiv \bar{Q}(\bar{q})$ and solving the condition $\bar{Q}(0) = 1$ for $p$ [40]. By doing this, we obtain

$$\bar{p}_c = \frac{\langle K\rangle}{\langle K^2\rangle - \langle K\rangle}. \qquad (10)$$

This is a well-known result for percolation on monoplex networks [41].

## III. COMPARISON OF THEORIES

We now examine several cases where the results we have derived for bond percolation on the edge-colored and projected versions of a multiplex network coincide.

### A. Uncorrelated layers

To begin with, let us consider some particular cases where the layers of the multiplex network are uncorrelated, i.e., where there is no correlation between the degrees of the different types of edges incident on each node.

*Case 1.*—For a two-layer multiplex network in which $\langle k_1 k_2\rangle = \langle k_1\rangle\langle k_2\rangle$, $\langle k_1^2\rangle < \infty$, and $\langle k_2^2\rangle < \infty$, the critical bond occupation probabilities of the edge-colored and projected networks are equal, $p_c = \bar{p}_c$, provided

$$\frac{\langle k_1^2\rangle - \langle k_1\rangle^2}{\langle k_1\rangle} = \frac{\langle k_2^2\rangle - \langle k_2\rangle^2}{\langle k_2\rangle}. \qquad (11)$$

Note that we are not imposing equal mean degrees in each layer. In the particular case where the multiplex consists of two Poisson random networks, Eq. (11) is trivially satisfied as both sides equal 1.

The derivation of Eq. (11) is as follows. The critical bond occupation probability $p_c$ in the edge-colored version of the multiplex is given by Eq. (7), where $\mu_{\max}$ is the positive root of $a\mu^2 + b\mu + c = 0$ with $a = -1$,

$$b = \frac{\langle k_1^2\rangle}{\langle k_1\rangle} + \frac{\langle k_2^2\rangle}{\langle k_2\rangle}, \qquad (12)$$

and

$$c = \frac{\langle k_1 k_2\rangle^2 - \langle k_1^2\rangle\langle k_2^2\rangle}{\langle k_1\rangle\langle k_2\rangle}. \qquad (13)$$

According to Eq. (10), the critical bond occupation probability in the monoplex projection of the multiplex is

$$\bar{p}_c = \frac{\langle k_1\rangle + \langle k_2\rangle}{\langle(k_1 + k_2)^2\rangle - \langle k_1\rangle - \langle k_2\rangle}. \qquad (14)$$

By setting $p_c = \bar{p}_c$ and then applying $\langle k_1 k_2\rangle = \langle k_1\rangle\langle k_2\rangle$, we recover Eq. (11).

*Case 2.*—For an $M$-layer multiplex network in which $\langle k_\alpha k_\beta\rangle = \langle k_\alpha\rangle\langle k_\beta\rangle$, $\langle k_\alpha\rangle = \langle k_\beta\rangle$, $\langle k_\alpha^2\rangle = \langle k_\beta^2\rangle$, and $\langle k_\alpha^2\rangle < \infty$ for all $\alpha, \beta = 1, \ldots, M$, we also have that the critical bond occupation probabilities of the edge-colored and projected networks are equal.

For the derivation of this property, we first define $z = \langle k_\alpha\rangle$ and $\sigma = \langle k_\alpha^2\rangle$ for all $\alpha = 1, \ldots, M$. Then, from Eq. (6), we obtain

$$\det(\mathbf{B} - \mu\mathbb{I}) = [f + Mz - \mu](f - \mu)^{M-1}, \qquad (15)$$

where $f \equiv f(\sigma, z) = (\sigma/z) - z$. This gives us $\mu_{\max} = f + Mz$, and therefore, from Eq. (7), we have

$$p_c = \frac{1}{f + Mz - 1}. \quad (16)$$

For the monoplex projection, we have $\bar{p}_c = 1/(f + Mz - 1)$ directly from Eq. (10); thus, $p_c = \bar{p}_c$.

### B. Correlated layers

Next, let us consider multiplex networks with positively correlated layers. There is a straightforward way to define a two-layer multiplex network with this attribute. Let $\rho_{k_1}$ and $\rho_{k_2}$ be the degree distributions of each layer, respectively. Then, the multidegree distribution $P_{k_1,k_2}$ of the multiplex network can be defined as

$$P_{k_1,k_2} = \nu \rho_{k_1} \delta_{k_1,k_2} + (1-\nu)\rho_{k_1}\rho_{k_2}, \quad (17)$$

where $\nu \in [0,1]$ is a parameter governing the correlation between the two layers. We can measure this correlation with the Pearson correlation coefficient $r$, which is related to $\nu$ by the following formula:

$$r = \nu \frac{\langle k_1^2 \rangle - \langle k_1 \rangle \langle k_2 \rangle}{\sqrt{\langle k_1^2 \rangle - \langle k_1 \rangle^2}\sqrt{\langle k_2^2 \rangle - \langle k_2 \rangle^2}}. \quad (18)$$

Note that each average in this formula is calculated on a single layer. Therefore, this expression can be used to create a multiplex network with a desired correlation between $k_1$ and $k_2$.

From the linearity of the multidegree distribution in Eq. (17), it is easy to see that if $p_c = \bar{p}_c$ in the uncorrelated case ($\nu = 0$), then the only possibility for $p_c$ and $\bar{p}_c$ to be different is the maximally correlated case ($\nu = 1$).

*Case 3.*—If we have an $M$-layer multiplex network, where (i) the layers of the multiplex are maximally correlated, $P_{\vec{k}} = \rho_{k_1}\delta_{k_1=k_2=\ldots=k_M}$, and (ii) $\langle k_\alpha^2 \rangle < \infty$ for all $\alpha = 1, \ldots, M$, then $p_c = \bar{p}_c$.

From condition (i), we have $\langle k_\alpha \rangle = \langle k_1 \rangle$ for all $\alpha = 1, \ldots, M$, so the elements of matrix **B** in Eq. (6) are all $B_{\alpha\beta} = \langle k_1^2 \rangle / \langle k_1 \rangle$. Therefore, $\mu_{\max} = M\langle k_1^2 \rangle / \langle k_1 \rangle$, and from Eq. (7), we obtain

$$p_c = \frac{1}{M\frac{\langle k_1^2 \rangle}{\langle k_1 \rangle} - 1}. \quad (19)$$

From Eq. (10), we directly have $\bar{p}_c = 1/(M\langle k_1^2 \rangle/\langle k_1 \rangle - 1)$; thus, $p_c = \bar{p}_c$.

So far, we have only looked at the value of the critical bond occupation probability in the edge-colored and projected networks. However, there is a remarkable result for multiplex networks made of layers that are Poisson random networks. To wit, it cannot only be proved that $p_c = \bar{p}_c$ but also that $S = \bar{S}$ for all values of $p \in [0,1]$.

*Case 4.*—Suppose we have a two-layer multiplex network, where each layer $\alpha \in \{1,2\}$ is defined by a Poisson degree distribution $\rho_{k_\alpha} = e^{-z_\alpha} z_\alpha^{k_\alpha}/k_\alpha$, with $z_\alpha = \langle k_\alpha \rangle$. Then, $S = \bar{S}$ for all $p \in [0,1]$ if (i) the layers of the network are uncorrelated, with $P_{k_1,k_2} = \rho_{k_1}\rho_{k_2}$, or (ii) the layers are maximally correlated, with $P_{k_1,k_2} = \rho_{k_1}\delta_{k_1,k_2}$.

We first consider the networks defined by condition (i). From Eq. (2), we have

$$q_\alpha = 1 - e^{-(z_1+z_2)pq_\alpha}, \quad (20)$$

for $\alpha \in \{1,2\}$. Similarly, from Eq. (8), we have

$$\bar{q} = 1 - e^{-(z_1+z_2)p\bar{q}}. \quad (21)$$

Thus, $q_1 = q_2 = \bar{q} = q$, and so solving either Eq. (20) or (21) for $q$ and substituting this solution into Eqs. (3) and (9), respectively, we obtain $S = \bar{S}$ for any $p$.

Next, applying condition (ii), Eq. (2) gives us

$$q_\alpha = 1 - (1 - pq_\alpha)e^{zpq_\alpha(pq_\alpha-2)}, \quad (22)$$

for $\alpha \in \{1,2\}$, and Eq. (8) gives us

$$\bar{q} = 1 - (1 - p\bar{q})e^{zp\bar{q}(p\bar{q}-2)}. \quad (23)$$

Thus, as for condition (i), we have $q_1 = q_2 = \bar{q} = q$, and in the same manner as before, we obtain $S = \bar{S}$ for any $p$.

From Case 4, we can also deduce that any linear combination of two Poisson layers with positive correlations yields $S = \bar{S}$ for all $p \in [0,1]$.

The results of this section have shown that, in many cases of interest, the predictions of the edge-colored and projected theories will coincide. We illustrate this fact in Fig. 1, where we compare the predictions of both theories for the value of $S$ against the results of numerical simulations of bond percolation on synthetic multiplex networks with two layers [Fig. 1(a)] and three layers [Fig. 1(b)]. In both figures, we include plots of the expected size of the second largest connected component (SLCC) to indicate the location of the percolation transitions [42]. In Fig. 1(a), the multiplex network consists of two Poisson random networks, one with mean degree $z = 3$ and the other with $z = 8$. The layers are maximally correlated ($\nu = 1$). In Fig. 1(b), the multiplex network consists of one Poisson random network with $z = 3$ and two scale-free networks, one with exponent $\gamma = 2.7$ and the other with $\gamma = 3$. The degree distribution for each of these layers is given by the following power law:

$$\rho_{k_\alpha} = \frac{k_\alpha^{-\gamma}}{\zeta(\gamma, k_\alpha^{\min})}, \quad (24)$$

where $\zeta(\gamma, k_\alpha^{\min}) = \sum_{i=0}^{\infty}(i + k_\alpha^{\min})^{-\gamma}$ is the Hurwitz zeta function and $k_\alpha^{\min}$ is the minimum degree in the layer [43]. We set $k_\alpha^{\min} = 1$ for both scale-free layers ($\alpha \in \{2,3\}$). All layers are uncorrelated ($\nu = 0$).

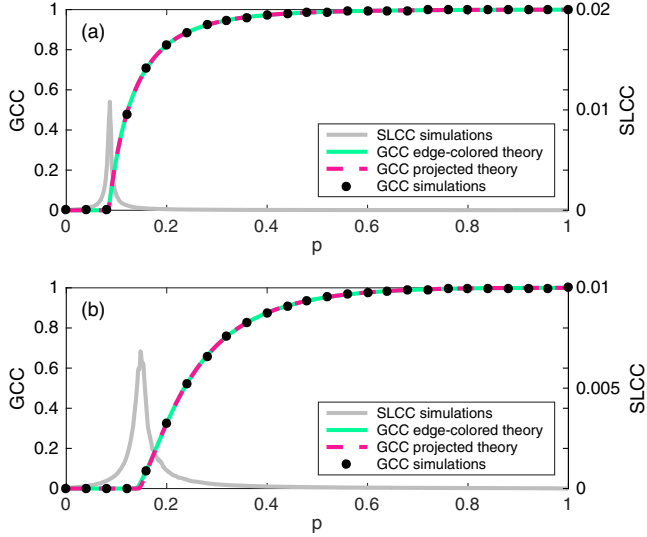

FIG. 1. Expected size of the GCC as a function of $p$ for (a) a maximally correlated two-layer multiplex network made of two Poisson random networks and (b) an uncorrelated three-layer multiplex network made of one Poisson random network and two scale-free networks. Numerical simulations are averaged over 100 realizations, and $N = 10^5$. Peaks in the expected size of the SLCC indicate percolation transitions.

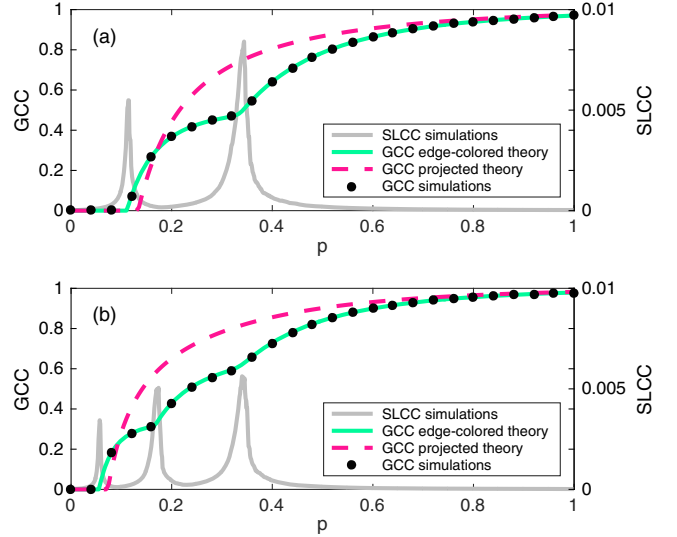

FIG. 2. Expected size of the GCC as a function of $p$ for (a) a two-layer multiplex network made of two Poisson random networks and (b) a three-layer multiplex network made of three Poisson random networks. Both networks are highly anticorrelated. Numerical simulations are averaged over 100 realizations, $N = 10^5$, and the overlap has 100 nodes. Peaks in the expected size of the SLCC indicate percolation transitions.

## IV. MULTIPLE PERCOLATION PHASE TRANSITIONS

The results of Sec. III are for multiplex networks with layers that are either uncorrelated or correlated with each other. In this section, we consider a class of multiplex networks with layers that are highly anticorrelated. In these networks, if a randomly chosen node is incident on an edge of a particular color, then it is highly unlikely that it is also incident on edges of any other color. We can construct such multiplex networks as follows. For simplicity, we describe only the two-layer case.

We begin with a maximally anticorrelated multiplex network with multidegree distribution

$$P_{k_1,k_2} = \frac{1}{2}\rho_{k_1}\delta_{k_2,0} + \frac{1}{2}\rho_{k_2}\delta_{k_1,0}. \quad (25)$$

This network consists of two completely separate layers, with half of the edges of type 1 and the other half of type 2. To allow the GCC to span both layers, we then connect them by allowing several nodes to be incident on edges of both types. We ensure that the relative number of these nodes is small in order to maintain high anticorrelation.

The results of bond percolation on this type of multiplex network are illustrated in Fig. 2. The multiplex network in Fig. 2(a) consists of two Poisson random networks, one with mean degree $z = 3$ and the other with $z = 9$. In Fig. 2(b), the multiplex network consists of three Poisson random networks with $z = 3$, $z = 6$, and $z = 18$, respectively. In both multiplex networks, only 100 nodes out of a total of $N = 10^5$ are incident on edges of every color. We refer to this subset of nodes as the *overlap* between layers [44]. Each of the remaining nodes is incident on edges of one color only. Therefore, both networks have highly anticorrelated layers. We see from Fig. 2 that these anticorrelated multiplex networks exhibit multiple percolation phase transitions and that the number of these transitions corresponds to the number of layers. The theory for bond percolation based on the monoplex projection of the multiplex network is no longer accurate, but our edge-colored theory of Sec. II accurately matches the numerical simulation results. The mechanism behind this phenomenon is that the most fragile (in terms of bond percolation) layer induces the degradation of the multiplex even though the other layer can still be perfectly operative, and because the way in which this happens is nonlinear, it cannot be described simply by the superposition of layers.

Note that if we relax the anticorrelation constraint, the above phenomenon can still be observed (see Fig. 3). In this example, the multiplex network consists of three Poisson random networks, each with $N = 10^4$ and $z = 50$, $z = 8$, and $z = 6$, respectively. The overlap between the first and second layers is $10^3$ nodes, and the overlap between the second and third layers is also $10^3$ nodes; however, these two overlapping sets are distinct. Although this multiplex can be viewed as being formed of five clearly differentiated subsets of nodes, Fig. 3 displays only a double percolation transition, which is perfectly captured by the edge-colored theory but not by the theory based on the monoplex projection of the multiplex.

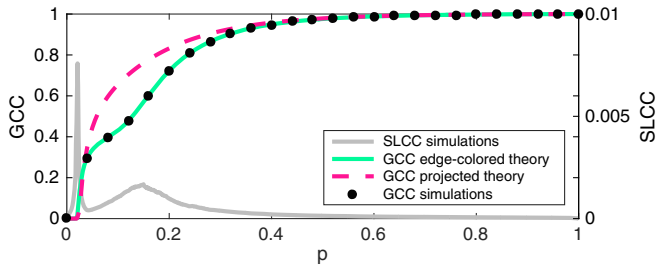

FIG. 3. Expected size of the GCC as a function of $p$ for a three-layer multiplex network made of three Poisson random networks. Each layer contains $N = 10^4$ nodes. Layers 1 and 2 have an overlap of $10^3$ nodes, as do layers 2 and 3. Peaks in the expected size of the SLCC indicate percolation transitions.

We posit that the reason we observe these multiple percolation transitions is as follows. When each layer of the multiplex network is designed to support a different dynamical process from those occurring on the other layers, as happens in multiplex transportation networks where layers can correspond, for example, to subway, road, train, etc., different connectivity patterns can be expected to emerge (e.g., different degree distributions, or the same degree distribution with different average degrees, etc.), and often these connectivity patterns will be anticorrelated (e.g., in geographic space, covering different areas). The emergence of multiple transitions is a consequence of these different connectivity patterns. This effect is similar to that observed in modular networks, where in essence, each module is separable from the other by deleting a few links, but not equivalent. The differences are subtle but important. In the case of multiplex networks, the construction of anticorrelated layers is responsible for the emergence of distinct structural patterns, not necessarily modules but with similar properties. In the multiplex case, there is a set of nodes that are common to both layers; this intersection forms a new structural pattern that can differ from the original distributions of each layer, and indeed it usually does.

## V. REAL-WORLD ANTICORRELATED MULTIPLEX NETWORKS

In this section, we provide an example of the phenomenon of multiple percolation phase transitions in a real-world anticorrelated multiplex network. The network in question is a two-layer multiplex constructed from the London rail transportation network [25] and the European air transportation network [36].

The data set used to construct the London rail transportation layer is itself a three-layer multiplex network of order 369, where nodes are train stations and edges are undirected routes between them. The three types of routes in this network are underground, overground, and DLR. To construct the first layer of our multiplex network, we aggregate the layers of this rail transportation network and make the resulting monoplex network unweighted. The data set used to construct the EU air transportation layer is a 37-layer multiplex network of order 450, where nodes are airports and edges are undirected routes between them. Each of the 37 different types of routes corresponds to a nonoverlapping subset of routes operated by a unique airline. Once again, to construct the second layer of our multiplex network, we aggregate the layers of this air transportation network and make the resulting monoplex network unweighted.

At this point in the construction of our multiplex network, the layers are completely separate. We connect the layers by observing that several of the stations in the London rail network are located at or within walking distance of the airports in the EU air transportation network. For example, there are five stops in the underground network at various terminals within Heathrow airport. If we define the walking distance as no more than 30 minutes, then there are 10 nodes that are incident on both rail routes and air routes. Taking account of this fact gives us a connected but still highly anticorrelated multiplex network.

In Fig. 4, we show the results of bond percolation on this network. It is clear from this figure that multiple percolation phase transitions are present. Note that as our approach is derived for multiplex networks with layers that are each internally unclustered, uncorrelated by degree, and of order $N \to \infty$, its accuracy in this case is affected by the network's complexity and finite size.

The significance of the network construction we have described is that it informs us of the ability of a traveler to traverse the London rail transportation system, reach a London airport, and connect to various destinations in Europe in the event of random failures within either layer. Reading Fig. 4 from right to left, we see that, as $p$ decreases, the depletion of the multiplex network is induced by the depletion of the rail transportation layer (which is the last layer to be incorporated into the GCC as the occupation probability $p$ increases). Findings like this could have novel implications for the European air transportation system, as it highlights the fact that the robustness of this system, at least with respect to connections between London and the rest of Europe, is influenced by the fragility of the rail network, which is localized entirely within London. Obviously, this result does not take into account other redundant connections like the bus transportation system or the road network; however, it is an illustration of how the interconnectivity of networked structures may suffer from fragilities induced by the most vulnerable layer.

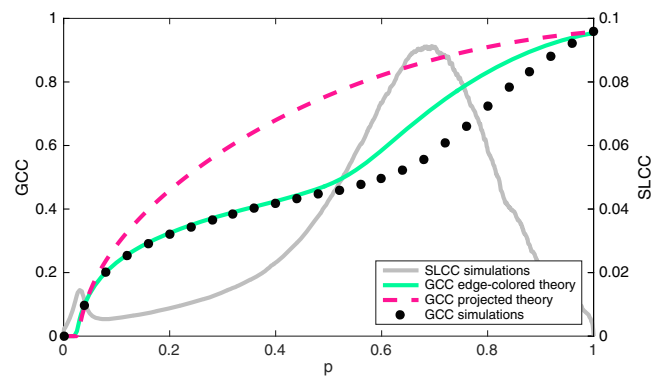

FIG. 4. Expected size of the GCC as a function of $p$ for a two-layer multiplex network constructed from the London rail transportation network and the EU air transportation network. Numerical simulations are averaged over 1000 realizations. Peaks in the expected size of the SLCC indicate percolation transitions.

From the evidence we have provided in this section and the previous one, we can expect that other anticorrelated multiplex networks, which may exist within real-world complex systems, may exhibit similar percolation properties.

## VI. CONCLUSION

In this paper, we have examined the process of bond percolation on multiplex networks. We have described an analytical approach to determine the expected size of the GCC and the value of the critical bond occupation probability. We have discussed why it is interesting to study these properties and addressed questions concerning the differences between bond percolation on a multiplex network and its monoplex projection. We have found that if the layers of the multiplex are uncorrelated or correlated with each other, then the bond percolation threshold can be the same in the monoplex projection as in the fully edge-colored multiplex. In fact, for multiplex networks composed of two layers that are uncorrelated or maximally correlated with each other and that both have Poisson degree distributions, the expected GCC size is the same in the edge-colored multiplex and its monoplex projection for all values of the bond occupation probability.

These results would seem to suggest that, for the process of bond percolation at least, the monoplex projection of a multiplex network tells us all we need to know about its phase-transition picture. However, when we considered multiplex networks with highly anticorrelated layers, we observed multiple percolation phase transitions. Our theory for edge-colored networks was able to capture this phenomenon, but the theory for the monoplex projection was no longer accurate. We have shown how multiple percolation transitions can occur in a real-world complex system, namely, a transportation system composed of the rail transportation network of London and the EU air transportation network. We anticipate that this phenomenon can be observed in other multiplex networks with anticorrelated layers across a diverse range of complex systems and scientific domains.

## ACKNOWLEDGMENTS

All authors were supported by the European Commission FET-Proactive Project PLEXMATH (Grant No. 317614). A. A. also acknowledges financial support from the ICREA Academia, Generalitat de Catalunya (2009-SGR-838), and the James S. McDonnell Foundation, and S. G. and A. A. were supported by FIS2012-38266. J. P. G. acknowledges funding from Science Foundation Ireland (Grants No. 11/PI/1026 and No. 09/SRC/E1780). D. C. also acknowledges funding from Science Foundation Ireland (Grant No. 14/IF/2461).